\documentclass[12pt]{article}
\usepackage{graphicx}

\title{Universality of Zipf's Law}
\author{Kenji Kawamura and Naomichi Hatano\\
{\it Department of Physics, Aoyama Gakuin University}\\
{\it Chitosedai, Setagaya, Tokyo 157-8572}}
\date{}

\begin{document}
\maketitle

\begin{abstract}
We introduce a simple and generic model that reproduces Zipf's law.
By regarding the time evolution of the model  as a random walk in the logarithmic scale, we explain theoretically why this model reproduces Zipf's law.
The explanation shows that the behavior of the model is very robust and universal. 

\end{abstract}

\section{Introduction}
Zipf's law is the observation that the frequency $P$ of the occurrence of various events is an inverse power-law function $P(>s)\propto s^{-1}$, where the rank $s$ is determined by the descending order of the frequency of each event.\cite{zipf}
Zipf first made this observation for the frequency of occurrence of English words in literature;
that is, the most frequent word \lq\lq the'' (rank $s=1$) appears twice as many times as the second most frequent word \lq \lq of''(rank $s=2$).
Since then, Zipf's law has been found in various fields, including the population of cities\cite{population} and the asset distribution of companies.\cite{takayasu,takayasu2}  
For example, Zipf's law for income distribution asserts that the size and number of companies are in inverse proportion; that is, the number of companies that have more than $10x$ assets is one-tenth of the number of companies that have more than $x$.
This universality of Zipf's law, however, has not been well explained theoretically.
To date, we have only models specific to each problem. \cite{population} 

In the present paper, we introduce a simple and generic model that reproduces Zipf's law.
We can regard this model both as the time evolution of the asset distribution and that of the population of cities.
We also explain theoretically why this model reproduces Zipf's law. 
Our explanation shows that Zipf's law of our model is very robust.

\section{Model and Simulation Results}
In this section we introduce our model and the results of its simulation. 
The model that we introduce here evolves as follows:
First, we consider a set of positive values $x_{i}$ with $N$ entities.
These values may be the assets of $N$ companies or the population of $N$ cities.
For simplicity, we assume that the initial values of $x_{i}$ are all equal.
Then we repeat the following procedures $T$ times:
\begin{enumerate}
\item Choose an entity $i$ randomly from $1\leq i\leq N$. 
\item For $i>1$, we move the amount $\alpha x_{i-1}$ from the $(i-1)$th entity to the $i$th entity, where $\alpha$ is a constant parameter with $0<\alpha <1$.
In other words
\begin{eqnarray}
x_{i-1}\rightarrow x_{i-1}-\alpha x_{i-1} \label{xm} \\
x_{i}\rightarrow x_{i}+\alpha x_{i-1}. \label{xp}
\end{eqnarray}
This procedure may be regarded as a deal between two companies or population movement between two cities.
For $i=1$, we increase all values $x_{i}$ by the same amount $x_{0}\alpha/N$.
\item Rearrange the entities in the descending order of the values $x_{i}$. 
\end{enumerate}

Let us explain this model from the viewpoint of asset distribution.
We focus on the hierarchical structure of the companies. 
Assume that a company will tend to trade with a company of similar size. 
That is, money flows from one company to a smaller company and the smaller company delivers products to the larger company. 
Hence, in our model, money flows from high-rank companies to low-rank companies as in Eqs.(\ref{xm}) and (\ref{xp}).

\begin{figure}
%\hskip -4cm
%\vskip -4cm
\includegraphics[width=0.5\textwidth]{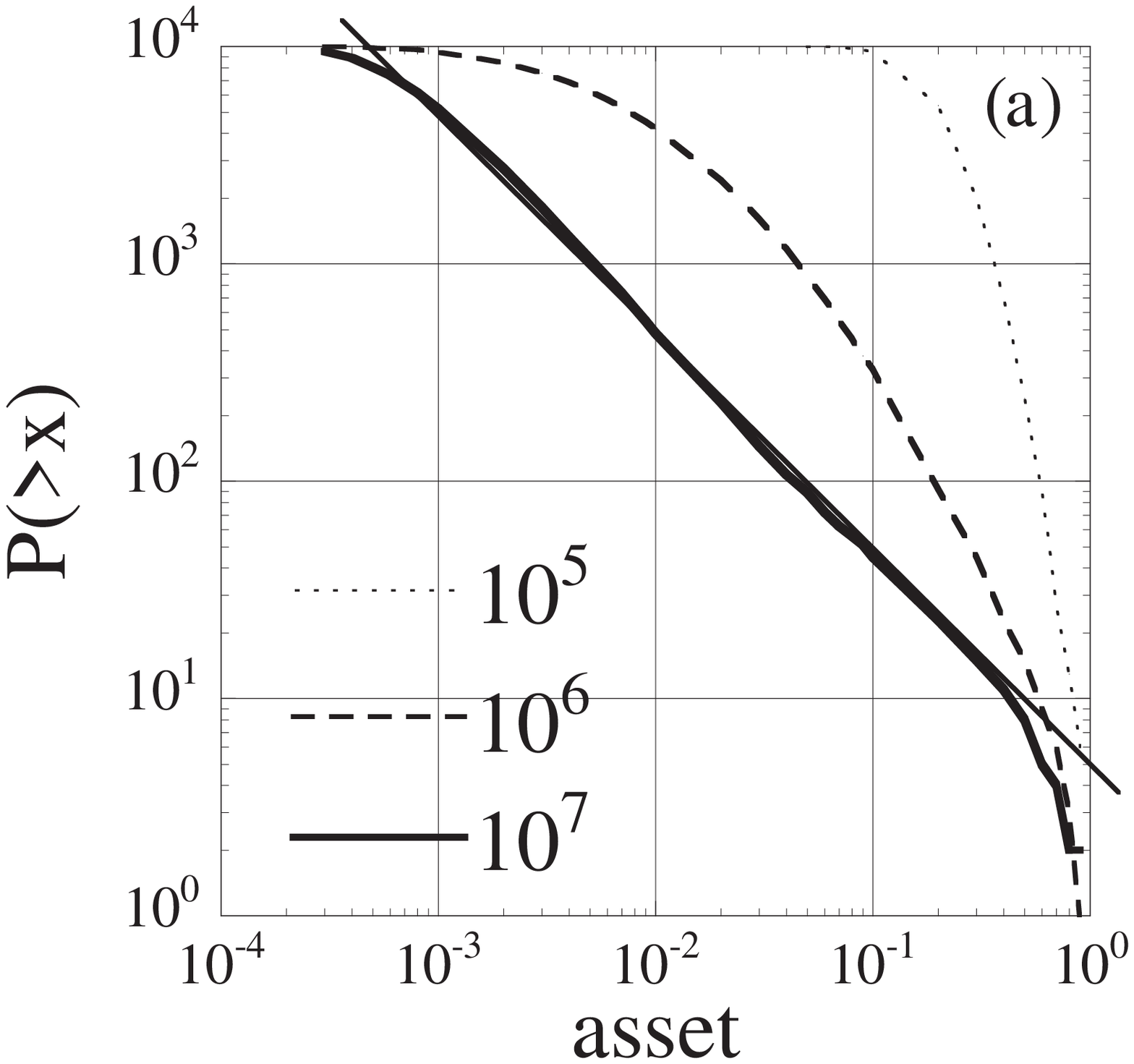}
\includegraphics[width=0.56\textwidth]{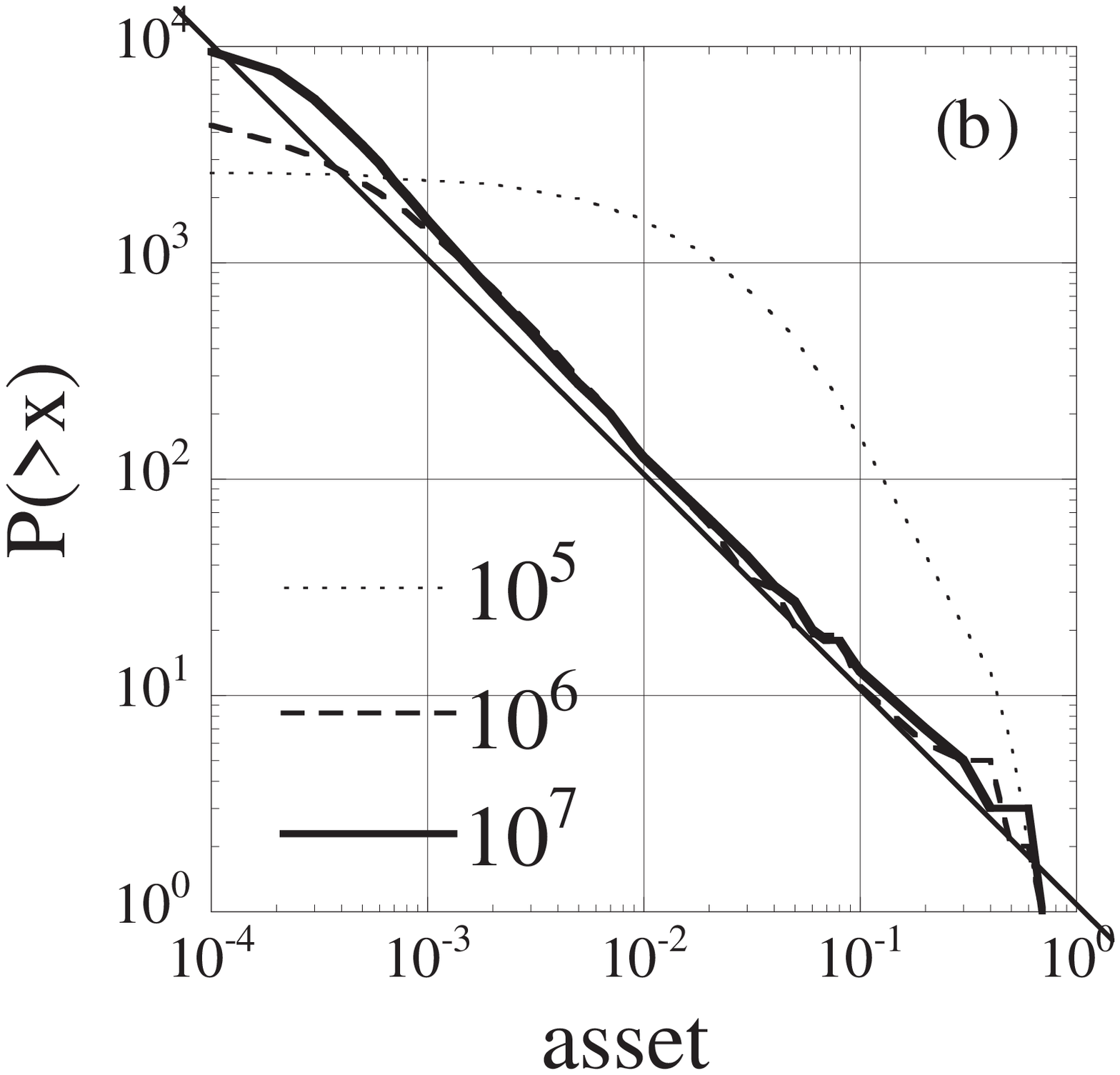}
\includegraphics[width=0.5\textwidth]{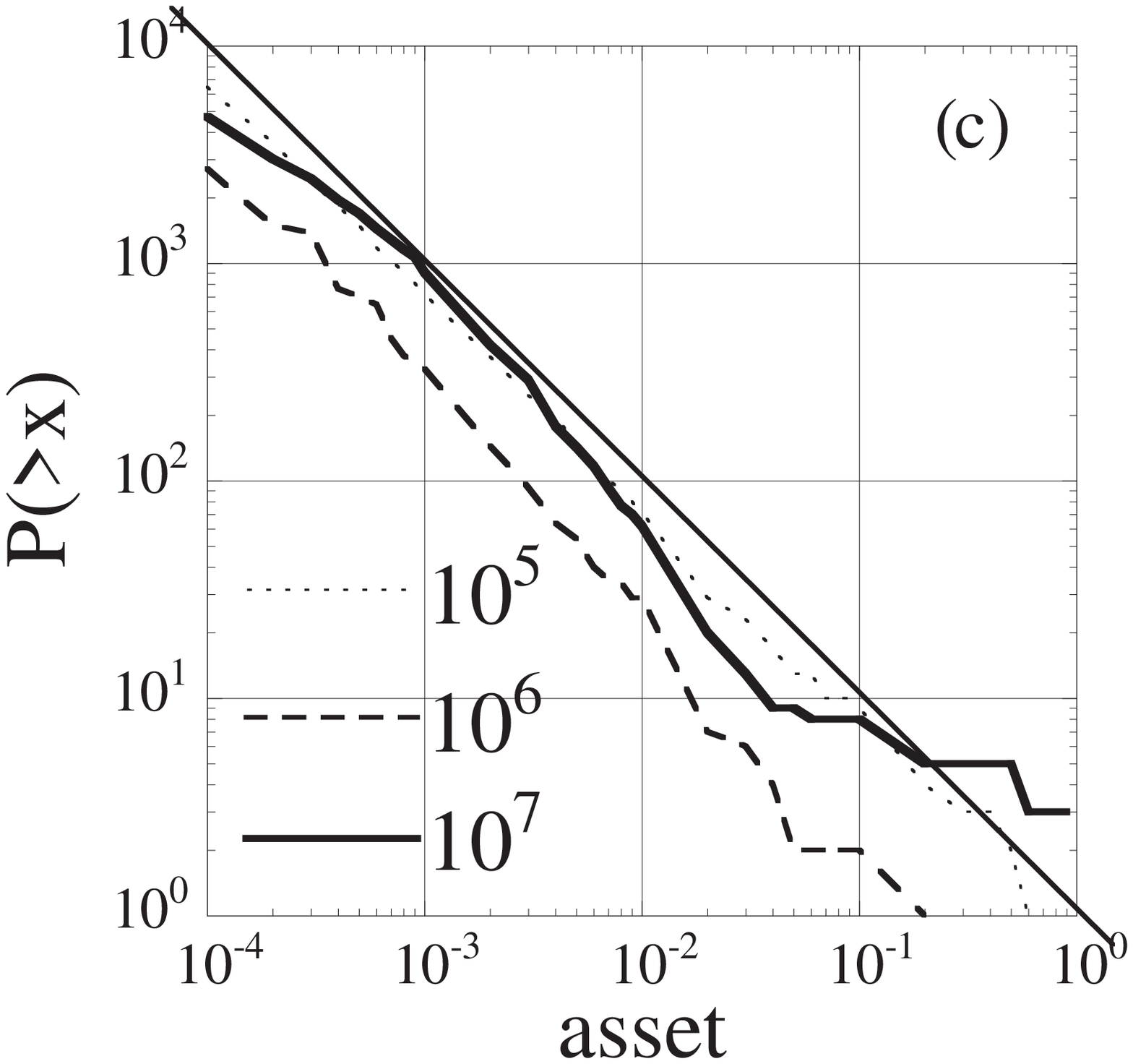}
\caption{The cumulative asset distribution of our model: 
(a) $\alpha=0.1$; (b) $\alpha=0.3$;  (c) $\alpha=0.99$. 
In each plot, the number of steps is $T=10^5$ (dotted line), $10^6$ (dashed line) and $10^7$(bold line). 
We normalized the results by the asset of the highest entity. 
The straight line indicates the power law with exponent $-1$, namely Zipf's law.}
\label{asset}
\end{figure} 

  Figure \ref{asset} shows results of the simulation of the above procedures for various values of $\alpha$. 
  Here the number of the companies is $N=10^4$.
In every case, the model reproduces Zipf's law as $T \rightarrow \infty$.
 We see that the parameter $\alpha$ is related to the rapidity of the convergence to Zipf's law.
In Fig.\ref{asset}(a) ($\alpha=0.1$) the distribution evolves slowly, but eventually converges to Zipf's law.

\section{Continuum Limit and Universality}
In this section we give a theoretical explanation as to why the model in the previous section reproduces Zipf's law.
Let us consider the limit $N \rightarrow \infty$.
Then, the value of the $i$th entity is nearly equal to the value of the $(i-1)$th entity, that is $x_{i-1} \approx x_{i}$.
Hence we regard the set of variables $x_{i}$ as a continuous variable $x$.
We consider in this limit the time evolution of the probability distribution function of $x$.
We note that Eqs.(\ref{xm}) and (\ref{xp}) give probability flows in the directions
\begin{eqnarray}
x \rightarrow x-\alpha x=(1-\alpha)x, \label{x-} \\ 
x \rightarrow x+\alpha x=(1+\alpha)x. \label{x+}
\end{eqnarray}
The evolution equation of $P(x)$ is therefore the following:
\begin{equation}\label{x}
\frac{\partial}{\partial t}P(x,t)=-2\gamma P(x,t)+\frac{\gamma}{1-\alpha}P(\frac{x}{1-\alpha},t)+\frac{\gamma}{1+\alpha}P(\frac{x}{1+\alpha},t),
\end{equation}
where $P(x,t)$ is the probability distribution of $x$ at time $t$ and $\gamma $ is a certain constant. 
The first term is a flow from the point $x$ to the points $(1\pm \alpha)x$, while the second and third terms are flows from $x/(1 \pm \alpha)$ into $x$.
The coefficients $1/(1 \pm \alpha)$ in front of $P$ in Eq.(\ref{x}) are necessary in order to satisfy the probability conservation.
To see this, we integrate Eq.(\ref{x}) over $x$ as
\begin{eqnarray}\label{int}
\frac{\partial}{\partial t} \int_0^\infty P(x,t)dx&=&-2\gamma \int_0^\infty P(x,t)dx 
+\frac{\gamma}{1-\alpha}\int_0^\infty P(\frac{x}{1-\alpha},t)dx \nonumber \\ 
&+&\frac{\gamma}{1+\alpha}\int_0^\infty P(\frac{x}{1+\alpha},t)dx.
\end{eqnarray}
Changing the variables to be
\begin{equation}
x' =\frac{x}{1-\alpha}, \qquad
x''=\frac{x}{1+\alpha},
\end{equation}
we can rewrite Eq.(\ref{int}) as
\begin{eqnarray}
\frac{\partial}{\partial t} \int_0^\infty P(x,t)dx&=&-2\gamma \int_0^\infty P(x,t)dx
+\gamma \int_0^\infty P(x',t)dx' \nonumber \\
&+&\gamma \int_0^\infty P(x'',t)dx'' \nonumber 
= 0.
\end{eqnarray}
Thus, the total probability is conserved.

To see Eq.(\ref{x}) from a different point of view, we change the variable to be $\xi=\log x$. 
The probability distribution function is transformed to
\begin{eqnarray}
P(x,t)dx=P(e^\xi,t)e^\xi d\xi
               \equiv \tilde P(\xi,t)d\xi.
\end{eqnarray}
In other words, we define
\begin{eqnarray}
\tilde P(\xi,t)=P(x,t)x
                          =P(e^\xi,t)e^\xi.
       \end{eqnarray}
By rewriting $P$ in terms of $\tilde P$ as
\begin{eqnarray}
P(x,t)&=&e^{-\xi} \tilde P(\xi,t)
          =\frac{1}{x} \tilde P(\log x,t) ,\\
P(\frac{x}{1 \mp \alpha},t)&=&\frac{1 \mp \alpha}{x} \tilde P(\log \frac{x}{1 \mp \alpha},t)
          =\frac{1 \mp \alpha}{x} \tilde P(\xi -\log (1 \mp \alpha),t),
\end{eqnarray}
we transform the evolution equation (\ref{x}) to 
\begin{eqnarray}\label{xi}
\frac{\partial}{\partial t}\tilde{P}(\xi,t)&=&-2\gamma\tilde P(\xi,t)+\gamma\tilde P(\xi+\beta_{+},t)+\gamma\tilde P(\xi-\beta_{-},t),  
\end{eqnarray}
where
\begin{eqnarray}
\beta_{\pm} &\equiv& \mp \log(1\mp \alpha)>0.
\end{eqnarray}

We can see the evolution equation (\ref{xi}) as a random walk from $\xi$ to ($\xi \pm \beta_{\pm}$), that is, a random walk with a fixed step size. 
Although we can solve Eq.(\ref{xi}) exactly, it is obvious even without solving it that the stationary solution is given by $\tilde{P}(\xi,t)=$constant as $t \rightarrow \infty$. 
Changing the variable back to $x=e^{\xi}$, we have $P(x,t)\propto x^{-1}$ as $t \rightarrow \infty$, or Zipf's law.

\begin{figure}
\hskip -2cm
\vskip 0.5cm
\includegraphics[width=0.6\textwidth]{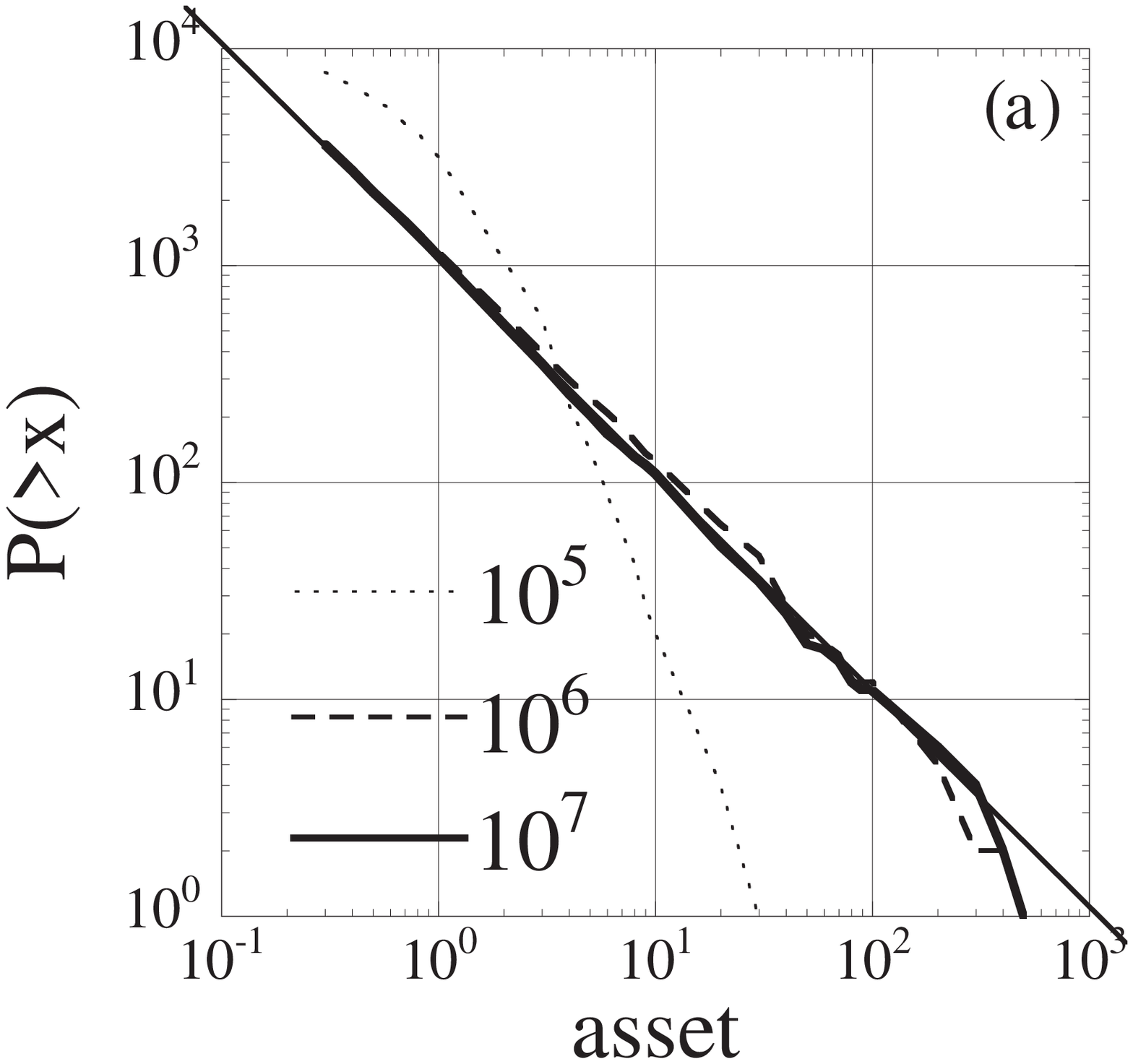}
\includegraphics[width=0.6\textwidth]{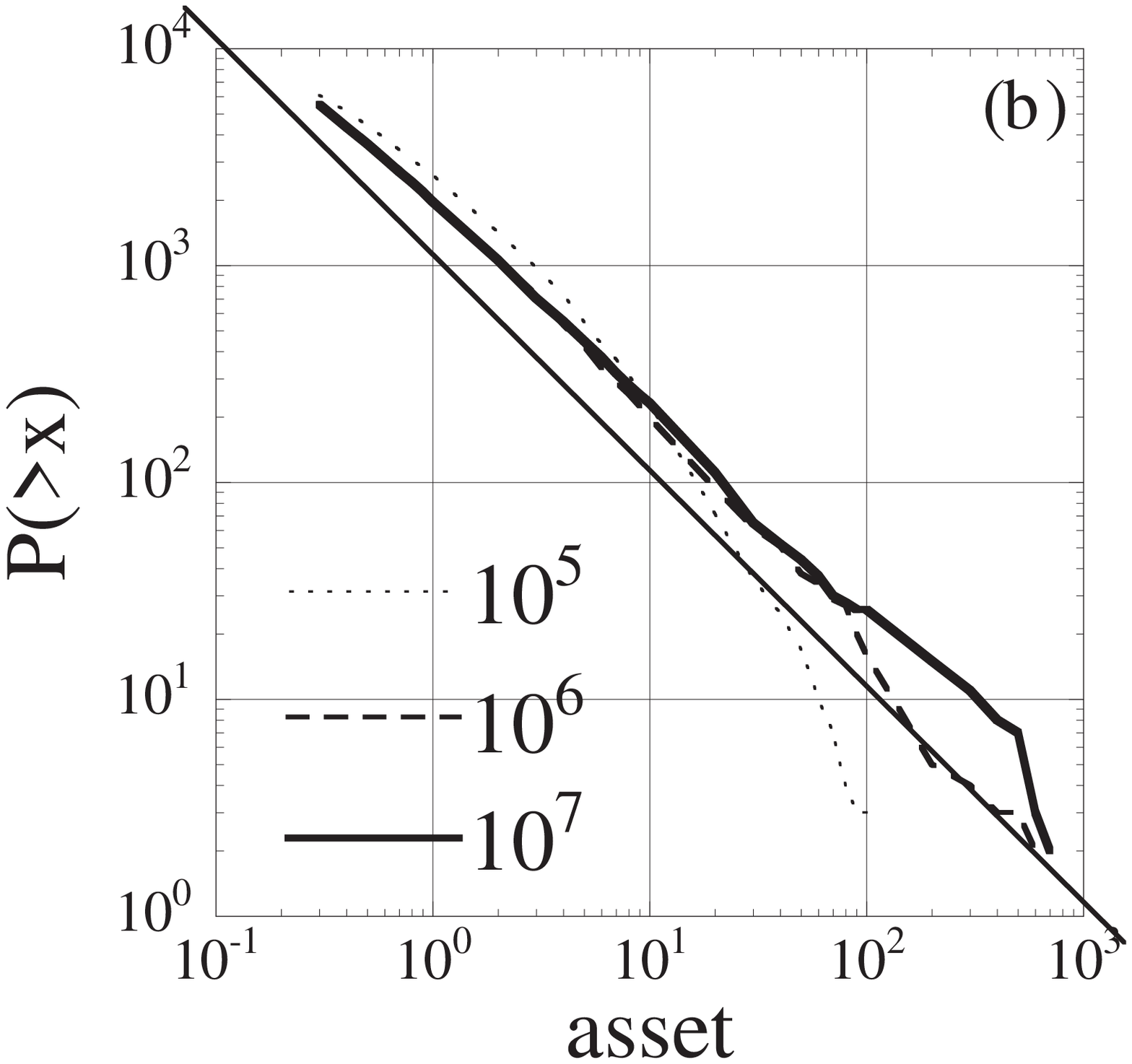}
\caption{The results of the simulations of the nonconservative model:
 (a) $\alpha=0.3$; (b) randomly chosen $\alpha$.
 In each plot, the number of steps is $T=$ $10^5$ (dotted line), $10^6$ (dashed line), $10^7$(bold line). 
We normalized the results by the asset of the highest entity. 
The straight line indicates the power law with exponent $-1$, Zipf's law.}\label{zipfex}
\end{figure}

From the above viewpoint, the resulting distribution always obeys Zipf's law as long as the diffusion of $x$ is uniform in the logarithmic scale.
We can thus generalize the model in $\S$2 extensively.
We assumed in $\S$2 that a chosen entity is given some value from the entity one rank higher in the model, but this is not necessary.
The model reproduces Zipf's law as long as there is a flow $\alpha x_{i}$ proportional to its size.
For example, the following procedure (\lq\lq the nonconservative model'') also reproduces Zipf's law:
\begin{enumerate}
\item Choose an entity $i$ randomly from $1\leq i\leq N$. 
\item Give or reduce randomly the amount $\alpha x_{i}$ from the chosen entity:
\begin{eqnarray}
x_{i} \rightarrow x_{i} \pm \alpha x_{i}.
\end{eqnarray}
\end{enumerate}
The simulation of the above procedure yields Fig.\ref{zipfex}(a).

It is  even unnecessary to make $\alpha$ a fixed value; Choosing $\alpha$ randomly from $0\leq \alpha \leq 0.99$ at each step in this nonconservative model, we obtain similar results to those shown in Fig.\ref{zipfex}(b).
 In this case, however, we need to bind the diffusion of $x$; otherwise, $P(x,t) \rightarrow 0 $ as $t \rightarrow \infty$.
 In Fig.\ref{zipfex}(b), we set the limit $10^{-4}\leq x\leq10^0$.
 
 \section{Summary}
In this paper, we explained the universality of Zipf's law theoretically and demonstrated it numerically.
The simulation results of our simple and generic model indicate that the present explanation of Zipf's law is applicable to various phenomena, both natural and social.
This paper does not consider the most famous case of Zipf's law, the frequency of English words.
We speculate, however, that Zipf's law of English words might be explained by the same time development as in this model.
 

\begin{thebibliography}{9}
 \bibitem{zipf}
 G. K. Zipf: \textit{Human Behavior and the Principle of Least Effort} (Addison-Wesley, Cambridge, 1949).
\bibitem{population}
For example, M. Marsili and Y.-C. Zhang: Phys. Rev. Lett. 80 (1998) 2741.
\bibitem{takayasu}
K. Okuyama, M. Takayasu and H. Takayasu: Physica A 269 (1999) 125.
\bibitem{takayasu2}
   H. Takayasu, M. Takayasu, M. P. Okazaki, K. Marumo and T. Shimizu: in 
  \textit{Paradigms of Complexity}, ed. M. M. Novak (World Scientific, Singapore 2000)  p.243.
\end{thebibliography}
\end{document}